\documentclass[iop,numberedappendix,appendixfloats,twocolappendix]{emulateapj}

\usepackage{hyperref}
\usepackage{epsfig}
\usepackage{natbib}
\usepackage{graphicx}    
\usepackage{makeidx}
\makeindex                       % To format bibliographies.
\usepackage{verbatim}                                   % Allows quoting source with commands.
\usepackage{amsmath,amsthm,amsfonts,amsopn,amssymb}     % Some nice math packages.m
\usepackage{longtable}
\usepackage{afterpage}
\usepackage{rotating} % Allows you to rotate figures, text and tables 
\usepackage{ulem}
\usepackage{epstopdf}  
\usepackage{multirow}	
\usepackage{color}
\usepackage[left]{lineno}
\usepackage{amsmath}
\usepackage{bm}
\usepackage{afterpage}
\usepackage{chngpage}

\newcommand{\rearth}{R$_{\oplus}$}
\newcommand{\mearth}{M$_{\oplus}$}

\newcommand{\degree}{^{\circ}}
\newcommand{\rprs}{R_{\rm{P}}/R_{*}}
\newcommand{\ars}{a/R_{*}}
\newcommand{\rp}{R_{\rm{P}}}
\newcommand{\massp}{M_{\rm{P}}}

\submitted{to ApJ}

\slugcomment{}
\shorttitle{A Search for Lost Planets}
\shortauthors{Schmitt et al. (2017)}

\begin{document}
\title{A Search for Lost Planets in the \textit{Kepler} Multi-planet Systems and the Discovery of the Long-period, Neptune-sized Exoplanet Kepler-150~\MakeLowercase{f}}
\author{
Joseph R. Schmitt\altaffilmark{1},
Jon M. Jenkins\altaffilmark{2}, 
Debra A. Fischer\altaffilmark{1}
}
\email{joseph.schmitt@yale.edu}

\altaffiltext{1}{Department of Astronomy, Yale University, New Haven, CT 06511 USA}
\altaffiltext{2}{NASA Ames Research Center, Moffett Field, CA 94035, USA}

\begin{abstract}
The vast majority of the 4700 confirmed planets and planet candidates discovered by the \textit{Kepler} mission were first found by the \textit{Kepler} pipeline. In the pipeline, after a transit signal is found, all data points associated with those transits are removed, creating a ``Swiss cheese''-like light curve full of holes, which is then used for subsequent transit searches.  These holes could render an additional planet undetectable (or ``lost'').  We examine a sample of 114 stars with $3+$ confirmed planets to evaluate the effect of this ``Swiss cheesing''.  A simulation determines that the probability that a transiting planet is lost due to the transit masking is low, but non-negligible, reaching a plateau at $\sim3.3\%$ lost in the period range of $P=400-500$ days. We then model all planet transits and subtract out the transit signals for each star, restoring the in-transit data points, and use the \textit{Kepler} pipeline to search the transit-subtracted (i.e., transit-cleaned) light curves. However, the pipeline did not discover any credible new transit signals. This demonstrates the validity and robustness of the \textit{Kepler} pipeline's choice to use transit masking over transit subtraction. However, a follow-up visual search through all the transit-subtracted data, which allows for easier visual identification of new transits, revealed the existence of a new, Neptune-sized exoplanet (Kepler-150~f) and a potential single transit of a likely false positive (Kepler-208).  Kepler-150~f ($P=637.2$ days, $\rp=3.64^{+0.52}_{-0.39}$ \rearth) is confirmed with $>99.998\%$ confidence using a combination of the planet multiplicity argument, a false positive probability analysis, and a transit duration analysis. 
\end{abstract}


\section{Introduction}
\label{sec:intro}

The \textit{Kepler} mission \citep{Borucki2010} has discovered more than 4700 \textit{Kepler} Objects of Interest (KOIs) that are classified either as confirmed planets \citep[$\sim2300$, e.g.,][]{Rowe2014, Morton2016} or planet candidates \citep[$\sim2400$, e.g.,][]{Coughlin2016}, according to the NASA Exoplanet Archive \citep{Akeson2013}.  The vast majority of these confirmed planets (CPs) and planet candidates (PCs) were initially discovered through the \textit{Kepler} pipeline. After calibrating the raw pixels, extracting the light curves, and correcting systematic errors, the Transiting Planet Search (TPS) and Data Validation (DV) modules search through the light curves for signatures of transiting planets, fitting limb-darkened transit models to the transit-like features, and constructing diagnostic tests assessing their planetary nature \citep{Jenkins2010a}. The potential transiting planet signatures are called Threshold Crossing Events (TCEs) and were reported throughout the primary and extended \textit{Kepler} mission \citep{Tenenbaum2012, Tenenbaum2013, Tenenbaum2014, Seader2015, Twicken2016}.  Follow-up vetting for potential PC status is then performed, the latest version of which is described thoroughly in Section 3 of \citet{Coughlin2016}. 

Additional vetting, such as follow-up observations, detailed light curve analysis, or statistical methods, can be used to confirm their planetary status or rule them out as (astrophysical) false positives (FPs), artifacts, or false alarms.  Many of these KOIs are in systems with multiple KOIs, which allows for easier statistical validation of their planetary status \citep{Lissauer2012, Lissauer2014, Sinukoff2016}.  As such, about half of the CPs discovered with \textit{Kepler} data are located in confirmed multiple planet systems.

A detail in the \textit{Kepler} planet search pipeline is its treatment of targets with multiple detections of TCEs. TPS passes the strongest TCE for each star to DV, which fits the TCE as a transit signature and performs the diagnostic tests. It then removes the TCE's in-transit data points from the light curve (filling them with noise) and calls the TPS algorithm to search for additional TCEs. Each additional TCE is subjected to the same light curve modeling and diagnostic tests. This iterative process continues until no more TCEs are found or until a maximum of nine are found. The removal of the in-transit data points creates what the \textit{Kepler} team calls a``Swiss cheese'' light curve \citep{Twicken2016}. This masking of data points can hide the existence of additional planets whose transits overlap with the previously found TCEs.

An additional wrinkle arises when trying to discover long-period planets in multiple planet systems.  Large, long-period planets with few transits are often best found by visually inspecting the light curves, especially for planets with $<3$ transits, even more so if they only transit once \citep{Wang2015, Uehara2016}. In fact, a citizen science program called Planet Hunters \citep{Fischer2012} that allows volunteers online to visually search the \textit{Kepler} data for exoplanets specializes in finding long-period planets. Through the power of visual inspection, Planet Hunters has led to the discovery of three exoplanets \citep{Schwamb2013, Wang2013, Schmitt2014b} and nearly 100 exoplanet candidates \citep{Lintott2013, Wang2015, Schmitt2014a, Schmitt2016}. However, while these planets might be easy to spot visually,  the fact that the light curve is filled with so many other planet transits can make them difficult to identify as new planets.  They can be easily mistaken for, or assumed to be, a transit from another, known planet in the system.  This problem could be fixed if the light curves had all known transit signals subtracted out, leaving only the previously undiscovered transits in the data.

In this paper, we examine the vast majority of the systems with three or more CPs and perform three separate analyses on them.  First, we simulate the percentage of true planets missed by TPS because of the algorithm's removal of known in-transit data points.  In our second test, we attempt to extract potential new planets in the data by subtracting out the transit signals of the known CPs and PCs rather than masking out their transits altogether, after which we then re-run TPS on the new, transit-subtracted light curves.  Finally, we then examine the transit-subtracted light curves visually to search for evidence of additional planets.

\section{Simulating for Lost Planets}
\label{sec:simulating}

\begin{figure*}
\includegraphics{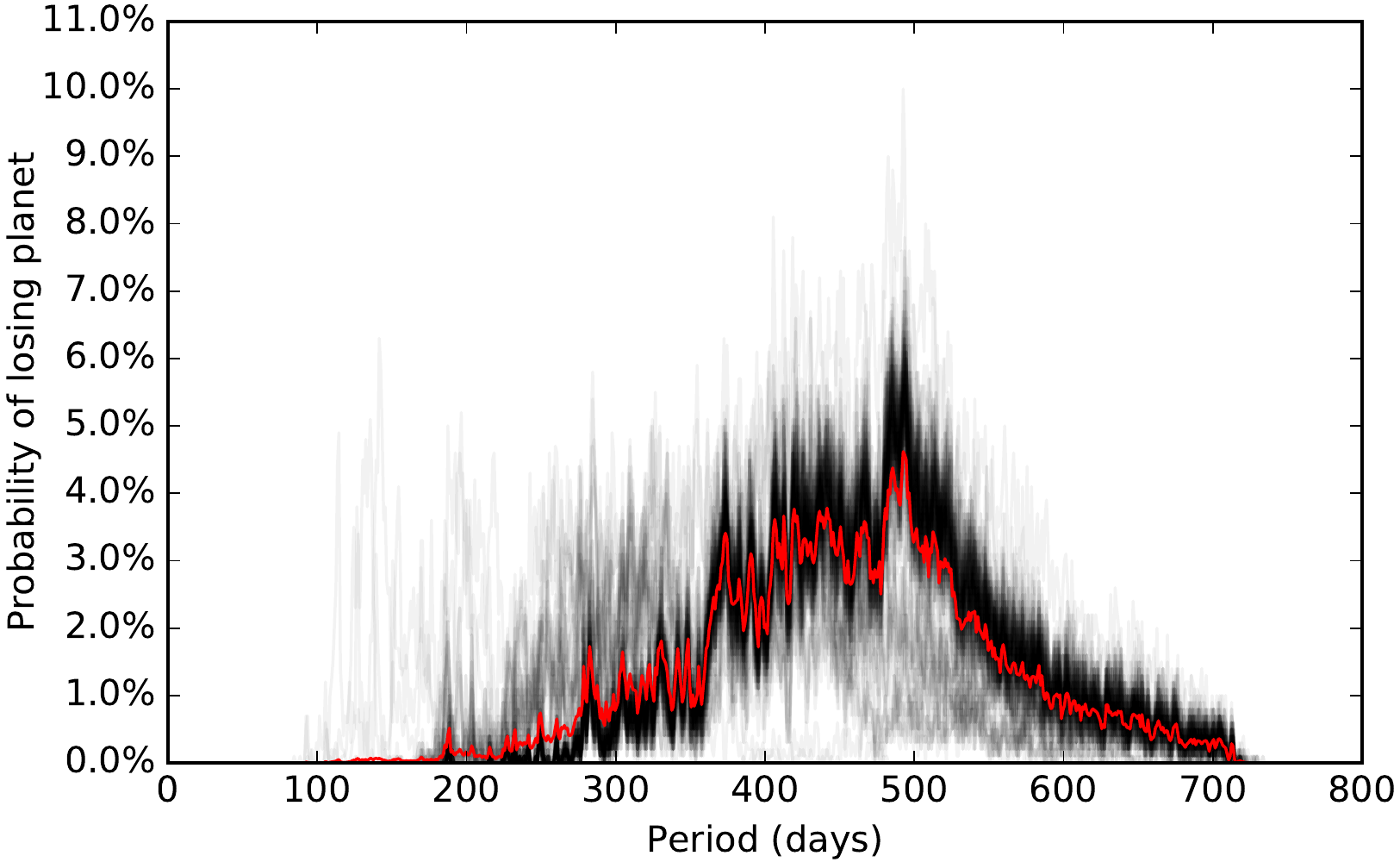}
\caption{Probability of a planet that was originally detectable (i.e., had $3+$ detectable transits) that became undetectable (i.e., had $<3$ detectable transits) after the in-transit data points of the successfully fit CPs and PCs were removed.  A transit was ruled ``detectable'' if at least 50\% of it was contained within the data (i.e., $<50\%$ in a data gap).  The probability of losing a planet at a certain period averaged over all stars is highlighted in red, while the star-by-star probabilities are shown in a transparent black, so that darker areas correspond to higher-density regions.}
\label{fig:lostplanets}
\end{figure*}

Flattening light curves, re-fitting for planets, and then re-running TPS is a time-intensive process.  We also might expect a higher rate of missed planets for systems with more known planets, as this corresponds to more data points being removed from subsequent transit searches (although higher multiplicity systems also attract additional manual scrutiny, which may or may not be the dominant effect). For these two reasons, we limited our study to only the 121 systems with three or more CPs, according to the NASA Exoplanet Archive \citep{Akeson2013} as of 2016 January 6. Some of these systems, however, were removed from the analysis in later steps (see Section~\ref{sec:searching}).  The final sample includes 114 stars (see Table~\ref{tab:stars} for the final list).

For each target star containing three or more CPs, we downloaded the long-cadence data from the Barbara A. Mikulski Archive for Space Telescopes (MAST).  In this sample of stars, the maximum time baseline was 1470 days.  For each of the 114 stars in the final sample, we then injected a planet into the light curve.  The planet's period and the star's mass and radius were used to calculate the transit duration (assuming inclination $i=0\degree$), and a random epoch was chosen. The simulated planet's transits were counted as detectable if at least 50\% of the transit was contained within the data (i.e., not in a data gap). The total number of transits detectable for each injection was then recorded. For this purpose, only the duration of the transit mattered, not the depth.  We then repeated this for many periods and epochs, injecting 1000 planets into each one day period bin from 2 to 1472 days uniformly distributed in period and phase (e.g., 1000 planets with a period between 2 and 3 days, 1000 planets with a period between 3 and 4 days, etc.).  A total of 169,050,000 planets were simulated.  

This procedure was then repeated (with the same simulated planets) on the light curve after all in-transit data points from known CPs and PCs were removed.  Only those in-transit data points that we were able to successfully remove in our subsequent analysis (see Section~\ref{sec:searching} and the Appendix) were removed in this step.  Therefore, a small number of CPs and PCs did not have their in-transit data points removed at this point.

The removal of the in-transit data points of the CPs and PCs resulted in some of the transits that were originally detectable (pre-removal) becoming undetectable.  This change in the window function (the Swiss cheesing of the light curve) generally would not be a problem when the planet transits the star many times.  However, TPS requires three transits to register a detection, and the loss of one or more transits may bring a planet that had $3+$ transits below that threshold.  These planets are then no longer detectable. These are the ``lost planets'', those that would have been detected if in-transit data points were properly corrected for instead of removed.

There are two shortcomings of this simulation.  One is that, for planets with five or more transits, the removal of $2+$ transits in a certain way could cause the subsequent planet detection to have an alias of the true period.  For example, removing the second and fourth transit in a system with five consecutive transits results in a detected period double that of the true period.  However, such events are rare, so its effects on the period detection are therefore ignored.  Another complication is that this simulation did not test to see how the transit signal-to-noise was affected, only how the number of detectable transits was changed.  Removing data points could reduce the signal-to-noise of undiscovered transits below the detection threshold.  This implies that we are underestimating the number of lost planets caused by transit masking. See Section~\ref{sec:discussion} for a more detailed discussion.

\begin{figure*}
\includegraphics{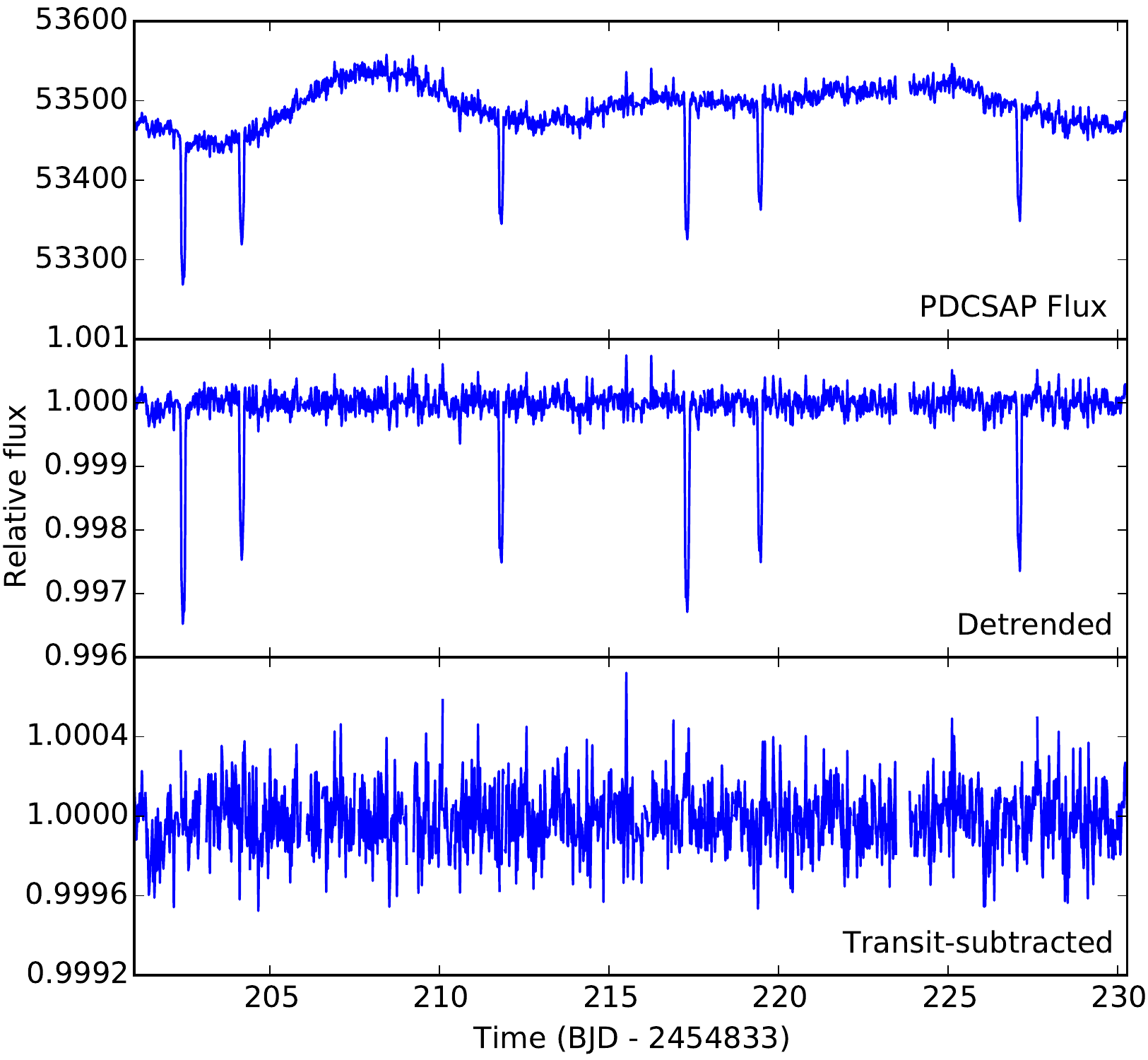}
\caption{Thirty day portion of the light curve for Kepler-18 (KIC 8644288).  \textit{Top:}  The PDCSAP flux, which was our starting point for the analysis.  \textit{Middle:}  The PDCSAP flux detrended for variability.  \textit{Bottom:}  The detrended flux after removing the transit signals from the CPs in Kepler-18. Note that the y-scaling changes in each panel.}
\label{fig:beforeandafter}
\end{figure*}

The probability of originally detectable planets becoming undetectable after the removal of the known in-transit data points is shown in Figure~\ref{fig:lostplanets}.  The red line is the probability averaged over all stars, while the transparent black lines in the background are the star-by-star probabilities, so that darker areas correspond to higher-density regions.  For the 114 stars in our sample, the lost planets are broadly distributed in period in a range of approximately 200-700 days.  (A small number of planets are lost below $P=200$~days, but this is almost exclusively for a small subset of stars that were observed for fewer quarters than the rest.)  The distribution plateaus between 400 and 500 days with a peak in the 480-500 day period range.  This 400-500 day peak has an average lost planet value of 3.3\%, meaning that 3.3\% of the observable, transiting planets in this region would be expected to be undetected (``lost'') after removing the in-transit points of previously found planets. In other words, if the planetary system had a planet in that period range, and if that planet transited, there would be a 3.3\% chance that it would not have been detectable due solely to the Swiss cheesing of the light curve.  The average's maximum of 4.6\% occurs at $P=493$~days.  The star-by-star peaks, on the other hand, vary between 3.8-10.0\%, with the peaks' locations ranging from $P=135$~days to $P=497$~days.  These numbers are likely to be lower limits because we have not taken into account the effect that removing some data points, but $<50\%$, would have on the detection statistics (see Section~\ref{sec:discussion} for a more thorough discussion of this).  These numbers, while small, are not negligible, and therefore imply that there may be a small number of missing exoplanets in the \textit{Kepler} data caused by the removal of in-transit data points of known planets.

\section{Searching for Lost Planets}
\label{sec:searching}

\begin{figure*}
\includegraphics{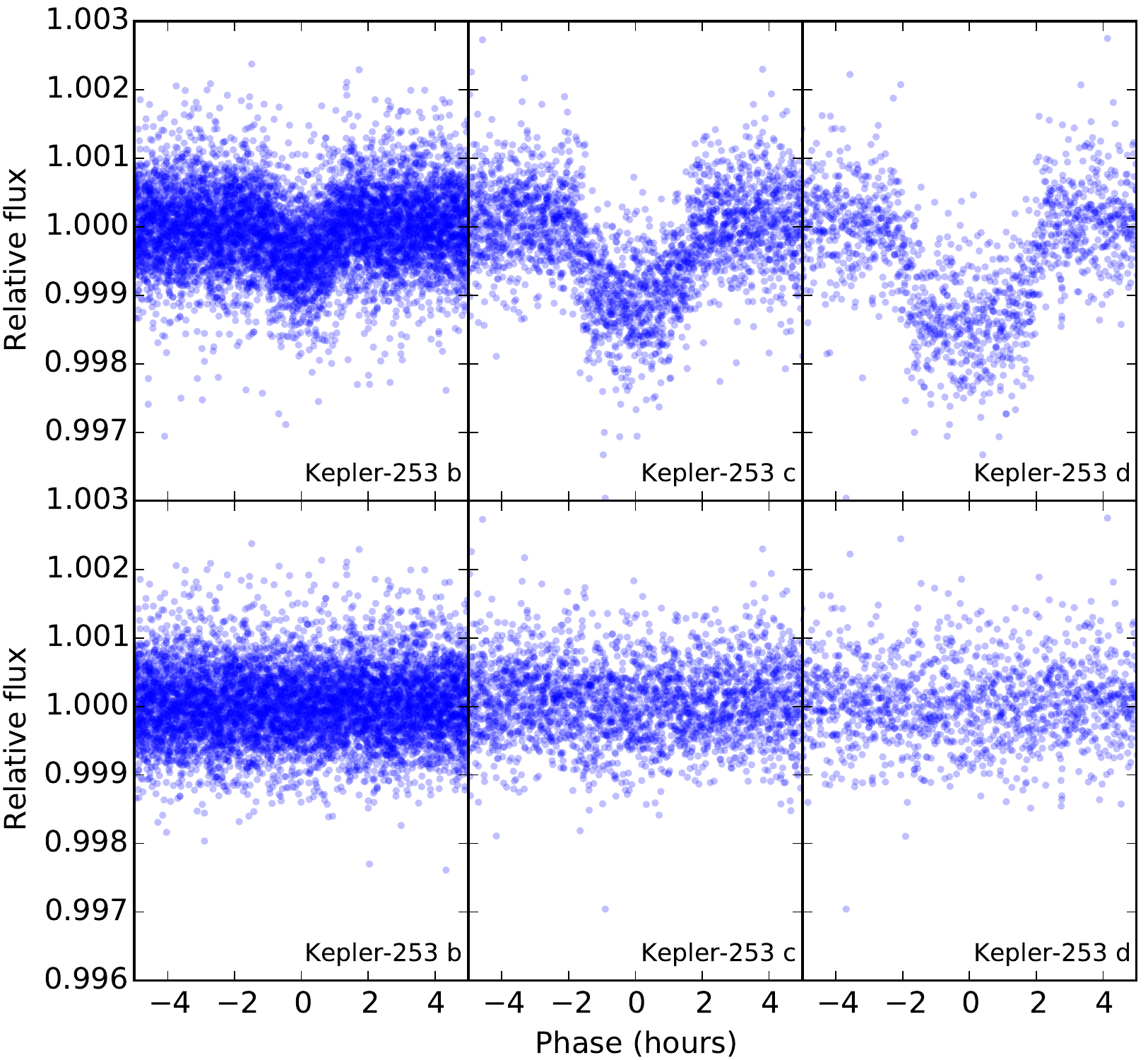}
\caption{Phase-folded light curve for all three CPs in Kepler-253. \textit{Top:} After detrending, but before transit subtraction.  \textit{Bottom:}  After detrending and transit subtraction. Data points are transparent to emphasize the transit shape.}
\label{fig:phasefold}
\end{figure*}

We began the analysis with the original Pre-search Data Conditioning Simple Aperture Photometry \citep[PDCSAP,][]{Stumpe2012, Smith2012, Stumpe2014} fluxes for every \textit{Kepler} star with $3+$ CPs.  We then removed the stellar variability using the PyKE \texttt{kepflatten} command in PyRAF \citep{Still2012}.  This command divides the light curve into chunks (or steps).  The step size is usually on the order of one to a few days.  It then fits the light curve in a window around (and including) the step with a polynomial (ignoring outliers).  The window size is usually approximately double that of the step size so that the edge effects of fitting do not affect the center portion of the window.  The window and step sizes were manually changed to fit each quarter of each star in order to remove as much of the stellar variability as possible without removing the transits.  These detrended light curves were then stitched together into one FITS file.  This was often successful as determined by eye (see Figure~\ref{fig:beforeandafter}).  However, a significant minority of stars had small, residual variability that could not be removed without disrupting the transit signal.  Attempting to completely fit the stellar variability would cause the transits to become partially filled in because the in-transit data points would not register as outliers.  The most egregious cases were those in which the frequency and magnitude of the stellar variability were greater than or approximately equal to the duration and depth of the planet, respectively. For these, the fitting procedure was unable to adequately fit the stellar variability without also removing the transit signal.  The worst cases were removed from the analysis. 

We then used the PyKE command \texttt{keptransit} \citep{Still2012} to fit the CPs and PCs in these systems across all observed quarters.  For each star, this was done iteratively starting with the CP or PC with the largest depth.  After it was successfully fit, its transits were subtracted from the detrended light curve.  Starting from this new, transit-subtracted and detrended light curve, we then performed the same procedure for the CP or PC with the next largest depth.  This was repeated until all CPs and PCs were fit and removed from the light curve.  Initial parameter guesses for the fits were taken from the KOI cumulative list, accessed 2016 January 8.  We chose to re-fit the transits rather than use the KOI list's values as fixed values to correct for any differences that our flattening might have induced in the transits.  A 30 day portion of this fitting procedure's results is shown for Kepler-18 in Figure~\ref{fig:beforeandafter}, and a complete, phase-folded version is shown for Kepler-253 in Figure~\ref{fig:phasefold}.

This fitting procedure was successful for the vast majority of cases, but not all.  If the majority of a star's CPs and PCs could not be successfully fit, it was removed from the analysis. Several examples of transit timing variations (TTVs) are also present in the data \citep{Mazeh2013, Holczer2016}.  The \texttt{keptransit} command is not equipped to handle TTVs, so these were not properly fit. CPs and PCs with TTVs from \citet{Mazeh2013} and \citet{Holczer2016} were noted, especially those which were visually apparent in the fitting results, in order to account for them in the later analysis.  In the most egregious cases, these TTVs were so large as to render the entire fit impossible or useless. For these reasons, Kepler-90, Kepler-247, and Kepler-279 were removed from this analysis.

The final count of systems that made it through all levels of analysis without being wholly removed is 114 stars, which host 397 CPs and 14 PCs. Of these, eight CPs and two PCs around nine stars were not successfully fit (see Table~\ref{tab:stars}).  These systems were still included in the analysis.

We then searched the detrended, transit-subtracted light curve for each of the 114 stars with TPS on the NASA Pleiades supercomputer.  TPS found 33 new, unique signals in 24 stars that did not correspond to known CPs and PCs.  The new signals had between three and six potential transits. Each of these signals were cross-checked with the locations of the removed transits of known planets. There were 13 new signals that overlapped with at least one transit that had been subtracted out.  Two of them overlapped with two removed transits, and four overlapped with three removed transits.  The other seven signals only overlapped with one removed transit.  The other 20 signals did not overlap at all with the removed transits and thus could potentially have been found in earlier TPS searches. Regardless, we examined all 33 signals more closely.

For each signal, we phase-folded the light curve according to its period and epoch.  Close visual examination of each signal revealed no credible transit-like signal.  After checking the original light curves, most were determined to be caused by edge effects 	from poorly corrected systematics in the original data. The other signals are spurious for undetermined reasons, but could potentially be attributed to improperly detrended light curves, poor transit subtraction, TTVs, or statistical noise.

\section{Visual Search for Planets}
\label{sec:visual}

Exoplanet systems with multiple transiting planets are likelier  than other systems to host more distant planets that also transit.  Some of these will only transit once or twice in the \textit{Kepler} data and are frequently missed by automated search algorithms.  While large planets with 1-2 transits can be easy to spot visually, these transits could easily be missed or overlooked in the forest of transits from known planets.  Light curves that are flattened, normalized, and have known transits subtracted out should then make these planets with 1-2 transits stand out more clearly.  Therefore, we visually inspected all the transit-subtracted light curves for additional transit signals. Three new potential transits were found around two stars.

\subsection{Kepler-150}
\label{subsec:kepler150}

Two highly significant transits were found in the light curve of Kepler-150 (KOI 408, KIC 5351250) and belong to a new planet, which we designate as Kepler-150~f.  The transits are visually apparent in both the PDCSAP flux and the raw SAP flux.  The second transit slightly overlaps with the transit of another planet in the system, but this was corrected for in the earlier transit subtraction.  A visual check confirms that there is no shorter period possible in the data.

\begin{figure}
\includegraphics{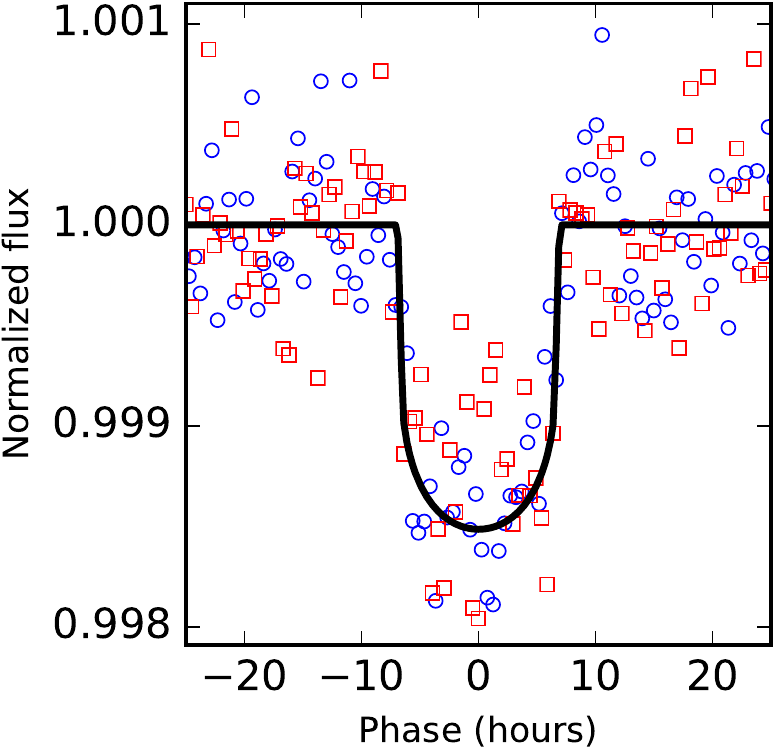}
\caption{Phase-folded, flux-normalized light curve for Kepler-150~f. Transits by other planets in this window were modeled and subtracted out.  Blue circles show the first transit, while red squares show the second.  The black line is the median best fit from \texttt{TAP}. }
\label{fig:kepler150f}
\end{figure}

The transits were fit with the IDL program \texttt{TAP} \citep{Gazak2012}, which is a Markov Chain Monte Carlo (MCMC) transit fitter that uses EXOFAST \citep{Eastman2013} to calculate transit models \citep{Mandel2002} using a wavelet-based likelihood function \citep{Carter2009}.  \texttt{TAP} fits for the basic transit parameters---the ratio of planet radius to stellar radius $\rprs$, the transit duration $T$, the impact parameter $b$, the midtransit times, and quadratic and linear limb darkening---in addition to white and red noise and a quadratic function to correct for improper normalization.  A circular orbit is assumed. Ten MCMC chains of length 200,000 were used to fit the transits.  The length was chosen so as to satisfy the Gelman-Rubin statistic \citep{Gelman1992} that tests for non-convergence.  The planet radius and semi-major axis were calculated by randomly drawing 1,000,000 values of $\rprs$ and $\ars$ from the posterior distributions of the \texttt{TAP} fit and 1,000,000 values of $R_{*}$ from the stellar replicated posterior distribution from the Q1-17 (DR25) stellar catalog \citep{Mathur2016}. An estimate of the planet's mass $\massp$ (and its associated error) was calculated using the mass-radius relationship from \citet{Weiss2014} using an additional normally distributed error with a standard deviation of 1~\mearth to account for the intrinsic variation in the relationship. The predicted radial velocity semi-amplitude was estimated using $P$, $\massp$, and $M_{*}$ from the stellar replicated posterior distribution in a similar manner as was done for the planet's radius and semi-major axis.  The reported best-fit values in Table~\ref{tab:tap} are the median values plus or minus $1\sigma$ error bars. The phase-folded, fitted light curve is shown in Figure~\ref{fig:kepler150f}.

Three arguments are used to confirm Kepler-150~f.  First, the fact that Kepler-150~f is found in a system with four other CPs argues strongly that it is not a FP \citep{Lissauer2012, Lissauer2014}.  According to \citet{Lissauer2012}, ``almost all of \textit{Kepler}'s multiple-planet candidates are planets''.  We used their Equation 6, shown below modified for systems with $4+$ CPs and/or PCs, to calculate the probability that Kepler-150~f is a FP.

\begin{equation}
P(\geq4\rm{\ planets\ +\ 1\ FP})=\frac{n_{4+}}{n_{\rm{t}}}\times\frac{n_{\rm{c}}(1-P)}{n_{\rm{t}}}\times n_{\rm{t}}
\end{equation}

\noindent where $n_{4+}$ is the number of systems with $4+$ CPs and/or PCs, $n_{\rm{c}}$ is the total number of CPs and PCs (where we adopt the same restriction as \citet{Lissauer2012} and require the planet radius to be $R_{\rm{P}}<22$~\rearth), $n_{\rm{t}}$ is the total number of stellar targets, and $P$ is the fraction of $n_{\rm{c}}$ that are true planets.  According to the KOI cumulative list accessed January 19, 2017, $n_{4+}=88$ and $n_{\rm{c}}=4208$.  We adopt the value of $n_{\rm{t}}=198,646$, as this is the number of stellar targets searched in \citet{Seader2015}, which was used in the latest KOI search \citep{Coughlin2016}.  This corresponds to 0.19-0.93 FPs in the systems with $4+$ KOI PCs for $P=0.9$ and $P=0.5$, respectively. Since there are 364 KOI PCs in systems with $4+$ KOI PCs, that gives a FP rate of 0.05-0.26\%, corresponding to a 99.74-99.95\% probability (3-3.5$\sigma$) that Kepler-150~f is a true planet.  

\begin{figure}
\includegraphics{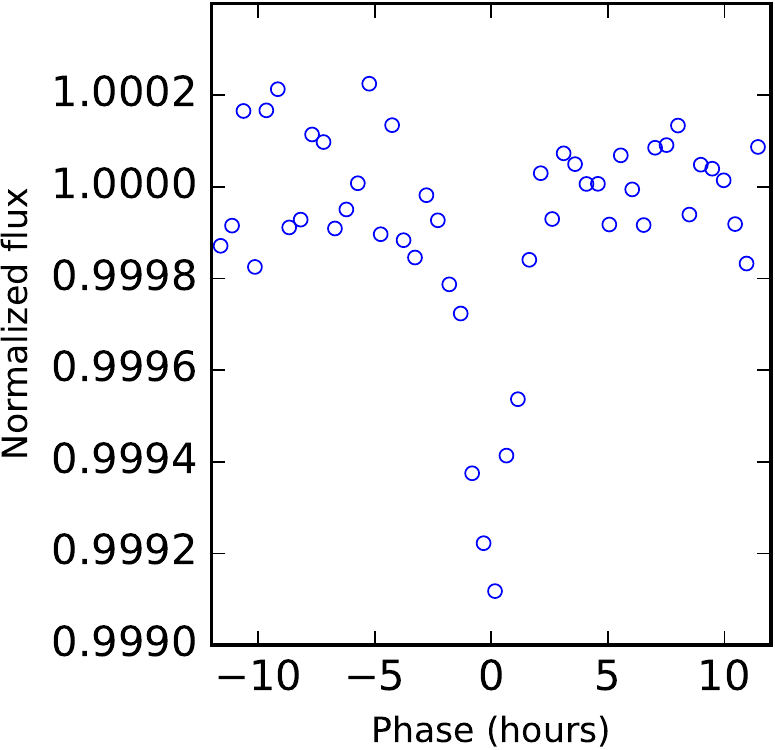}
\caption{Potential single transit in the light curve of Kepler-208 at 786.7641 BKJD. Due to its morphology and only having one transit, we did not attempt a transit fit.}
\label{fig:Vtransit}
\end{figure}

Secondly, we performed an analysis with the Python program \texttt{vespa} \citep{Morton2012, Morton2015, Morton2016}, which calculates a false positive probability (FPP) for three scenarios of FPs: an eclipsing binary, a hierarchical eclipsing binary, and a background eclipsing binary. The \texttt{vespa} analysis of Kepler-150~f results in $\rm{FPP}=0.69\%$.  This does not taken into account the planet multiplicity argument.

\begin{deluxetable*}{lcc}
\tablewidth{0pt}
\tablecaption{Kepler-150 f properties.}
\tablehead{
\colhead{Parameter}       & 
\colhead{Best-fit Value}  &
\colhead{Unit}            } 
\startdata
Period ($P$)            & $637.2093^{+0.0169}_{-0.0154}$ & days \\ 
Impact parameter ($b$)  & $0.00^{+0.77}_{-0.76}$                \\ 
Inclination ($i$)       & $90.00\pm0.19$                 & deg  \\ 
Duration ($T$)          & $13.41^{+0.59}_{-0.38}$       & hours \\ 
Planet radius to stellar radius ratio ($\rprs$) & $0.0358^{+0.0041}_{-0.0022}$ \\
Semi-major axis to stellar radius ratio ($\ars$) & $291.1^{+62.9}_{-106.5}$ \\
First midtransit time   & $509.0334^{+0.0125}_{-0.0148}$ & BKJD  \\ 
Second midtransit time  & $1146.2422^{+0.0082}_{-0.0077}$ & BKJD \\ 
Planet radius ($\rp$)   & $3.64^{+0.52}_{-0.39}$       & \rearth \\ 
Semi-major axis ($a$)   & $1.24^{+0.29}_{-0.45}$         & AU    \\
False positive probability (FPP) & $<0.0018$ & \%                \\   
Kepler-band             & 14.985                         & mag   \\ 
Mass estimate ($\massp$)  & $9.01^{+1.37}_{-1.51}$     & \mearth \\
Radial velocity semi-amplitude estimate & $0.72^{+0.11}_{-0.12}$ & m/s 
\enddata
\tablecomments{Best-fit results for Kepler-150~f from the \texttt{TAP} transit fit and its derived parameters.  Reported values are the median and the upper and lower $1\sigma$ error bars.  Eccentricity was held fixed at zero.}
\label{tab:tap}
\end{deluxetable*}

Lastly, all planets in the Kepler-150 system were tested to see if their periods and transit durations were consistent with orbiting the same star.  This transit duration analysis has been used previously as an additional level of vetting \citep{Steffen2010, Lissauer2012, Chaplin2013, Cabrera2014}.  In a perfectly coplanar, circular, edge-on system, and assuming that the planets' radii and masses are much smaller than that of the star's, the transit duration $T$ and the period $P$ of the planets are, for any pair of planets, related according to the formula $T_{i}/P_{i}^{1/3} = T_{j}/P_{j}^{1/3}$, where the $i$ and $j$ indices refer to any two planets in the system.  The values for $T_{b,c,d,e}$ and $P_{b,c,d,e}$ were taken from the KOI catalog, while $T_f$ and $P_f$ were determined by \texttt{TAP}.  The $T/P^{1/3}$ values are highly consistent with each other with no planet being $>1.6\%$ different from the average value. This strongly indicates that all five planets orbit the same star and also suggests that they have nearly circular orbits.

These three arguments should all be independent of each other.  Therefore, one can multiply their FPPs together to get a combined FPP.  Assuming the conservative FP rate of 50\% from the multiplicity argument and combining the multiplicity argument's FPP with the \texttt{vespa} FPP leads to a combined $\rm{FPP}=0.0018\%$.  Including the duration analysis implies a value smaller than this.  Therefore, we conclude with $>99.998\%$ ($4.3\sigma$) certainty that Kepler-150~f is a true exoplanet.  It is approximately the size of Neptune with a planet radius $\rp=3.64^{+0.52}_{-0.39}$~\rearth and a period $P=637.2093^{+0.0169}_{-0.0154}$~days.

\subsection{Kepler-208}
\label{subsec:Kepler208}

Two potential single transits were discovered in Kepler-208 (KOI 671, KIC 7040629), a system with four confirmed planets.  However, one transit has been previously discovered at Barycentric Kepler Julian Date\footnote{To convert to Barycentric Julian Date (BJD), use the formula $\rm{BJD}=\rm{BKJD}+2,454,833.0$} $\rm{BKJD}=786.7641$ by \citet{Uehara2016}, who performed their own visual checks of KOI systems.  The other transit is highly suspect.  The potential new transit is sharply V-shaped (see Figure~\ref{fig:Vtransit}) and has a short duration of just $T=2.65$~hours.  Its morphology is most consistent with a single transit of a background eclipsing binary, although there is a slight possibility that this could be a large, distant planet in a glancing transit. However, because of its morphology and only having a single transit, we performed no further analysis of this transit.

\section{Discussion}
\label{sec:discussion}

It was not unexpected that ``lost'' planets caused by the removal of known in-transit data points would be rare. There are two reasons why masking out data may result in losing planets.  One reason is that it could reduce the number of detectable transits from $3+$ to $<3$.  However, the parameter space for which this would occur is small to begin with.  Planets with periods around 400-500 days (for stars with a full $\sim1470$ day baseline) are the most susceptible to be lost because these are likely to transit exactly three times.  Due to geometry, planets with $P=400-500$~days are unlikely to transit in the first place.  If such a planet were to transit though, then even a single overlapping transit could remove it from detectability. However, because long-period planets typically have longer durations, it is possible for the signal to persist, even when a few data points from the transit are lost because they are superimposed on a short-period transit.  Therefore, if they overlap with a short-period planet, which is the most likely overlapping scenario, then only a small portion of the long-period planet's transit is removed, which would generally allow it to still be detectable.  In order to remove enough of the long-period transit to render it undetectable, either a) it would need to overlap with another fairly long-period planet, b) two or more short-period planets would need to overlap the same long-period transit in different spots, or c) a short-period planet and a data gap would need to overlap the same long-period transit in different spots.  Each of these requires an unlikely confluence of events. 

A minor confounding factor would be the relative impact parameters of the planets with overlapping transits.  Since planetary systems are usually flat with small scatter, inner planets are more likely to have lower impact parameters than outer planets in the system \citep{Fang2012, Fabrycky2014}.  Higher impact parameters result in shorter durations. Therefore, it is possible, although unlikely, that the transit duration of the longer-period planet would be shorter than (or, at least, comparable to) that of a shorter-period planet in the same system.  This would make overlapping a larger portion of the long-period planet's transit easier and thus could more easily render the long-period planet undetectable. 

A second way that masking out data may result in losing planets is for weak transit signals that were on the verge of detectability in the first place.  The Multiple Event Statistic (MES) is the signal-to-noise of a transit signal in the \textit{Kepler} pipeline \citep{Jenkins2002}.  TPS requires a minimum MES of $7.1\sigma$ for detection and classification as a TCE \citep{Jenkins2010b, Tenenbaum2012}.  Removing a full transit or even small portions of a transit could result in the MES dropping below this threshold, thus rendering the planet lost.  For example, a four-transit $8\sigma$ signature becomes $6.9\sigma$ when one whole transit is removed, dropping it below the detection threshold.  Subtracting out previously found transits rather than removing the data points altogether may keep the $\rm{MES}>7.1\sigma$. On one hand, this might be hard to promote to  PC or CP status anyway since the FP population is dominated by TCEs with three transits and a low MES \citep{Mullally2015}.  On the other hand, the fact that these systems already have $3+$ CPs imply that these three-transit, low MES cases could be more easily proven to be planets through statistical validation \citep{Lissauer2012}.  Note that our simulation of planets in Section~\ref{sec:simulating} did not test for changes in MES directly.

A visual search of the data, however, resulted in the discovery of Kepler-150~f.  This makes Kepler-150 just one of 26 stars to host at least five exoplanets, according to the NASA Exoplanet Archive \citep{Akeson2013}.  The demise of the main \textit{Kepler} mission, however, has made additional follow-up study of this system difficult.  Detecting the stellar reflex motion through radial velocities is challenging with current instruments given the $\approx0.7$~m/s Doppler semi-amplitude expected for a $\approx9.0$~\mearth planet.

The result of a single discovery of a new, long-period planet among these does not solve an outstanding question of just where the long-period planets are \citep[e.g.,][]{Lissauer2014}.  Only 10\% of the CPs and PCs in this analysis had $P>45$~days. Only nine planets (including Kepler-150~f) had $P>100$~days and just three had $P>200$~days. Kepler-150~f is the only planet with $P>300$~days in this sample of 114 planetary systems with $3+$ CPs.  The geometric probability to transit obviously plays a role in this.  The probability that a planet on a circular orbit transits its host star is simply $R_{*}/a$. On average, using their listed best-fit $a/R_{*}$ from the KOI list, the CPs and PCs in our sample with $P<100$~days are calculated to have transit probabilities $\sim10$ times higher than those with $P>100$~days (including Kepler-150~f), although there are about 45 times more shorter-period planets than longer-period planets. This is despite the fact that the average $\rp$ of the short- and long-period planets are approximately equal.	 Trying to solve this missing long-period planet question with long-period, eccentric orbits would only make this problem a little worse.  Eccentric planets are \textit{more} likely to transit than circular planets \citep{Burke2008}.  However, other potential solutions still exist, including differences in inclination between inner and outer planets and/or significantly different planetary architectures \citep{Moriarty2016}, or there could just simply be greater than expected detection issues.

\section{Conclusion}
\label{sec:conclusion}

A simulation of millions of planets around 114 stars with $3+$ confirmed planets showed that there is a low, but non-negligible, probability of transiting planets being lost after transit masking of known CPs and PCs (about 3.3\% of transiting planets in the period range of $P=400-500$ days).  We searched these same stars for new planetary transits with TPS, but instead of masking out transits of known CPs and PCs, we fit and subtracted them out. However, our search discovered no credible new transit signals, which was consistent with our simulations.

The original purpose of masking out known in-transit data points rather than subtracting them out was due to the fact that the subtraction process produced a large number of FPs due to improper subtraction.  This made it impractical to do in a time-intensive analysis such as the \textit{Kepler} pipeline despite the risk that it would cause planets to be missed. Our analysis, however, demonstrates the validity and robustness of the \textit{Kepler} pipeline's choice to use transit masking over transit subtraction.

However, a visual follow-up of the transit-subtracted light curves revealed the existence of a Neptune-sized exoplanet, Kepler-150~f ($\rp=3.64^{+0.52}_{-0.39}$~\rearth) with $>99.998\%$ confidence, making Kepler-150 one of the few stars with $5+$ known planets.  Because of its long period ($P=637.2$~days), only two transits are contained in the data, which made it undetectable to the \textit{Kepler} pipeline.  We attribute its discovery to the subtraction of known planet transits from the light curve.  This discovery suggests the possibility that improved light curve flattening and transit subtraction, or simply better eyes, may result in the discovery of new, long-period exoplanets.

\vspace{0.25in}\noindent\textit{Acknowledgements} 

D.F. and J.S. acknowledge NASA 14-ADAP14-0245.  This paper includes data collected by the \textit{Kepler} mission. Funding for the \textit{Kepler} mission is provided by the NASA Science Mission Directorate. We gratefully acknowledge the entire \textit{Kepler} mission team's efforts in obtaining and providing the light curves used in this analysis.  We also thank Bruce Clarke for his help in running the TPS tasks on the NAS Pleiades supercomputer.  Support for MAST for non-HST data is provided by the NASA Office of Space Science via grant NNX13AC07G and by other grants and contracts.  This research has made use of NASA's Astrophysics Data System Bibliographic Services.  This work made use of PyKE \citep{Still2012}, a software package for the reduction and analysis of Kepler data. This open source software project is developed and distributed by the NASA Kepler Guest Observer Office. 
 
\bibliographystyle{apj}
\bibliography{bib}

\appendix
\label{appendix}

\setcounter{table}{0}
\renewcommand{\thetable}{A\arabic{table}}

Not all planets were successfully removed from the analysis.  Table~\ref{tab:stars} lists the 114 stars in our final sample, the number of CPs and PCs in each star, and how many and which CPs and PCs were not successfully fit.  

{\LongTables
\begin{deluxetable*}{lrccc}
\tablewidth{0pt}
\tablecaption{Stars with $3+$ confirmed Used in This Study}
\tablehead{
\colhead{\textit{Kepler}}             & 
\colhead{KIC}                         & 
\colhead{Number of confirmed }        & 
\colhead{Number of KOI}               & 
\colhead{Name of non-fitted}          \\ 
\colhead{name}                        & 
\colhead{}                            &
\colhead{planets (not fit)}           & 
\colhead{candidates (not fit)}        &
\colhead{planets and KOIs}            }
\startdata
Kepler-11  &  6541920  &  6  &  0  &    \\
Kepler-18  &  8644288  &  3  &  0  &    \\
Kepler-20  &  6850504  &  5  &  0  &    \\
Kepler-30  &  3832474  &  3  &  0  &    \\
Kepler-31  &  9347899  &  3  &  1  &    \\
Kepler-33  &  9458613  &  5  &  0  &    \\
Kepler-37  &  8478994  &  3  &  0  &    \\
Kepler-42  &  8561063  &  3  &  0  &    \\
Kepler-48  &  5735762  &  3  &  0  &    \\
Kepler-49  &  5364071  &  4  &  0  &    \\
Kepler-52  &  11754553  &  3  &  0  &    \\
Kepler-53  &  5358241  &  3  &  0  &    \\
Kepler-54  &  7455287  &  3  &  0  &    \\
Kepler-55  &  8150320  &  5 (1)  &  0  &  Kepler-55 c  \\
Kepler-58  &  4077526  &  3  &  1  &    \\
Kepler-60  &  6768394  &  3  &  0  &    \\
Kepler-62  &  9002278  &  5  &  0  &    \\
Kepler-65  &  5866724  &  3  &  0  &    \\
Kepler-79  &  8394721  &  4  &  0  &    \\
Kepler-80  &  4852528  &  4  &  1  &    \\
Kepler-81  &  7287995  &  3  &  0  &    \\
Kepler-82  &  7366258  &  4  &  0  &    \\
Kepler-83  &  7870390  &  3  &  0  &    \\
Kepler-84  &  5301750  &  5  &  0  &    \\
Kepler-85  &  8950568  &  4  &  0  &    \\
Kepler-89  &  6462863  &  4  &  0  &    \\
Kepler-102  &  10187017  &  5  &  0  &    \\
Kepler-104  &  6678383  &  3  &  0  &    \\
Kepler-107  &  10875245  &  4  &  0  &    \\
Kepler-114  &  10925104  &  3  &  0  &    \\
Kepler-122  &  4833421  &  5  &  0  &    \\
Kepler-124  &  11288051  &  3  &  0  &    \\
Kepler-127  &  9451706  &  3  &  0  &    \\
Kepler-130  &  5088536  &  3  &  0  &    \\
Kepler-132  &  6021275  &  3  &  1  &    \\
Kepler-138  &  7603200  &  3  &  0  &    \\
Kepler-142  &  10982872  &  3  &  0  &    \\
Kepler-149  &  3217264  &  3  &  0  &    \\
Kepler-150  &  5351250  &  4  &  0  &    \\
Kepler-164  &  10460984  &  3  &  1 (1)  &  KOI 474.03  \\
Kepler-169  &  5689351  &  5  &  0  &    \\
Kepler-171  &  6381846  &  3  &  0  &    \\
Kepler-172  &  6422155  &  4  &  0  &    \\
Kepler-174  &  8017703  &  3  &  0  &    \\
Kepler-176  &  8037145  &  3  &  1  &    \\
Kepler-178  &  9941859  &  3  &  0  &    \\
Kepler-184  &  7445445  &  3  &  0  &    \\
Kepler-186  &  8120608  &  5  &  0  &    \\
Kepler-194  &  10600261  &  3  &  0  &    \\
Kepler-197  &  12068975  &  4  &  0  &    \\
Kepler-203  &  6062088  &  3  &  0  &    \\
Kepler-206  &  6442340  &  3  &  0  &    \\
Kepler-207  &  6685609  &  3  &  0  &    \\
Kepler-208  &  7040629  &  4  &  0  &    \\
Kepler-215  &  8962094  &  4  &  0  &    \\
Kepler-219  &  9884104  &  3  &  0  &    \\
Kepler-220  &  9950612  &  4  &  0  &    \\
Kepler-221  &  9963524  &  4  &  0  &    \\
Kepler-222  &  10002866  &  3  &  0  &    \\
Kepler-223  &  10227020  &  4 (1)  &  0  &  Kepler-223 e  \\
Kepler-224  &  10271806  &  4  &  0  &    \\
Kepler-226  &  10601284  &  3  &  0  &    \\
Kepler-228  &  10872983  &  3  &  0  &    \\
Kepler-229  &  10910878  &  3  &  0  &    \\
Kepler-235  &  4139816  &  4  &  0  &    \\
Kepler-238  &  5436502  &  5  &  0  &    \\
Kepler-244  &  6849310  &  3  &  0  &    \\
Kepler-245  &  6948054  &  3 (2)  &  1  &  Kepler-245 b, Kepler-245 d  \\
Kepler-249  &  7907423  &  3  &  0  &    \\
Kepler-250  &  8226994  &  3  &  0  &    \\
Kepler-251  &  8247638  &  4  &  0  &    \\
Kepler-253  &  8689373  &  3  &  0  &    \\
Kepler-254  &  9334289  &  3  &  0  &    \\
Kepler-256  &  9466668  &  4  &  0  &    \\
Kepler-257  &  9480189  &  3  &  0  &    \\
Kepler-265  &  5956342  &  4  &  0  &    \\
Kepler-267  &  10166274  &  3  &  0  &    \\
Kepler-272  &  10426656  &  3  &  0  &    \\
Kepler-275  &  3447722  &  3  &  1 (1)  &  KOI 1198.04  \\
Kepler-276  &  3962243  &  3  &  0  &    \\
Kepler-282  &  8609450  &  4  &  0  &    \\
Kepler-286  &  10858691  &  4  &  0  &    \\
Kepler-288  &  4455231  &  3  &  0  &    \\
Kepler-292  &  6962977  &  5  &  0  &    \\
Kepler-295  &  9006449  &  3 (1)  &  0  &  Kepler-295 d  \\
Kepler-296  &  11497958  &  5 (1)  &  0  &  Kepler-296 e  \\
Kepler-298  &  11176127  &  3  &  0  &    \\
Kepler-299  &  11014932  &  4  &  0  &    \\
Kepler-301  &  11389771  &  3  &  0  &    \\
Kepler-304  &  5371776  &  3  &  1  &    \\
Kepler-305  &  5219234  &  3  &  1  &    \\
Kepler-306  &  5438099  &  4  &  0  &    \\
Kepler-310  &  10004738  &  3  &  0  &    \\
Kepler-325  &  9471268  &  3  &  0  &    \\
Kepler-327  &  8167996  &  3  &  0  &    \\
Kepler-331  &  4263293  &  3  &  0  &    \\
Kepler-332  &  10328393  &  3  &  0  &    \\
Kepler-334  &  10130039  &  3  &  0  &    \\
Kepler-336  &  6037581  &  3  &  0  &    \\
Kepler-338  &  5511081  &  4  &  0  &    \\
Kepler-339  &  10978763  &  3  &  0  &    \\
Kepler-341  &  7747425  &  4  &  0  &    \\
Kepler-342  &  9892816  &  3 (1)  &  1  &    \\
Kepler-350  &  4636578  &  3  &  0  &    \\
Kepler-354  &  6026438  &  3  &  0  &    \\
Kepler-357  &  8164257  &  3 (1)  &  0  &  Kepler-357 d  \\
Kepler-363  &  6021193  &  3  &  0  &    \\
Kepler-372  &  11401767  &  3  &  0  &    \\
Kepler-374  &  6871071  &  3  &  2  &    \\
Kepler-399  &  5480640  &  3  &  0  &    \\
Kepler-402  &  7673192  &  4  &  1  &    \\
Kepler-444  &  6278762  &  5  &  0  &    \\
Kepler-445  &  9730163  &  3  &  0  &    \\
Kepler-446  &  8733898  &  3  &  0  &    
\enddata
\tablecomments{Stars included in this study, each with $3+$ CPs.  Some also have additional PCs.  The column ``Number of confirmed planets (not fit)'' refers to the number of CPs in that system, while the number in the parentheses, if applicable, are how many of those CPs were unable to be fit.  The column ``Number of KOI candidates (not fit)'' is similar, but for PCs that have not been confirmed.  The names of the non-fitted CPs and PCs are in the last column.}
\label{tab:stars}
\end{deluxetable*}
}

\end{document}